\documentclass[10pt,letterpaper]{article}
\usepackage[top=0.85in,left=2.75in,footskip=0.75in]{geometry}

\usepackage{array,epsfig,rotating}
\usepackage{multirow}
\usepackage{subfig}
\usepackage{graphicx}
\usepackage{xcolor}
\usepackage{amsfonts}
\usepackage{enumerate}
\usepackage{verbatim}
\usepackage{fancyhdr}
\usepackage{booktabs}
\usepackage[flushleft]{threeparttable}

\def\bfbeta{\boldsymbol \beta}
\def\rT{\mathrm T}

\newcommand{\bB}{{\bf B}}
\newcommand{\bbeta}{{\boldsymbol \beta}}
\newcommand{\bgamma}{{\boldsymbol \gamma}}

\usepackage{amsmath,amssymb}

\usepackage{changepage}

\usepackage[utf8x]{inputenc}

\usepackage{textcomp,marvosym}

\usepackage{cite}

\usepackage{nameref,hyperref}

\usepackage[right]{lineno}

\usepackage{microtype}
\DisableLigatures[f]{encoding = *, family = * }

\usepackage{array}

\newcolumntype{+}{!{\vrule width 2pt}}

\newlength\savedwidth

\usepackage[aboveskip=1pt,labelfont=bf,labelsep=period,justification=raggedright,singlelinecheck=off]{caption}

\makeatletter
\renewcommand{\@biblabel}[1]{\quad#1.}
\makeatother

\usepackage{lastpage,fancyhdr,graphicx}
\usepackage{epstopdf}
\fancyheadoffset[L]{2.25in}
\fancyfootoffset[L]{2.25in}
\begin{document}
\vspace*{0.2in}

\begin{flushleft}
{\Large
\textbf\newline{
{Estimation of time-varying reproduction numbers underlying epidemiological processes: a new statistical tool for the COVID-19 pandemic} 
 } 
}


\vspace{.5cm}

Hyokyoung G. Hong\textsuperscript{1} and 
Yi Li\textsuperscript{2*},
\\
\bigskip
\textbf{1} Department of Statistics and Probability,  Michigan State University, East Lansing, MI, USA \\
\textbf{2} Department of Biostatistics, University of Michigan, Ann Arbor, MI, USA  
\bigskip

%
%





* yili@umich.edu

\end{flushleft}
\section*{Abstract}

The coronavirus pandemic has rapidly evolved into an unprecedented crisis.
The susceptible-infectious-removed (SIR) model and its   variants  {have been} used for modeling the pandemic.
However, time-independent  parameters in the classical models may not capture the dynamic transmission and removal processes, governed by virus containment strategies taken at various phases of the epidemic. Moreover, few models account for possible inaccuracies of the reported cases.  We propose a Poisson model with time-dependent   
  transmission and removal rates to account for  possible random errors in reporting and  estimate a time-dependent disease  reproduction number, which may reflect the effectiveness of  virus control strategies.
  We apply our method to study the pandemic in  several severely impacted countries, and analyze and forecast the evolving spread of the coronavirus. We have developed an interactive web application to facilitate readers' use of our method.




\section{Introduction}
 {Coronaviruses  are enveloped single-stranded positive-sense RNA viruses belonging to a broad family of coronaviridae and are widely harbored in animals \cite{fehr2015coronaviruses,li2005bats,woo2012discovery}.}  Most of the coronaviruses only cause  mild respiratory infections, but SARS-CoV-2, a newly identified  member of the coronavirus family, initiated the  contagious and lethal coronavirus disease 2019 (COVID-19) in December 2019 \cite{trombetta2016human,wang2020evolving}.  Since the detection of the first case in Wuhan, the COVID-19 pandemic has evolved into a global crisis within only four months. As of {{June 30}, 2020}, the virus has infected more than {10} million individuals, caused about {518,000} deaths \cite{jhu}, and altered the life of billions of people. 
 
 The pandemic has been closely monitored by the international society. {For example,  the World Health Organization (WHO) and Johns Hopkins University's Coronavirus Resource Center \cite{jhu} have,  since the outbreak, reported the daily numbers of infectious and recovered cases, and deaths for nearly every country. The governmental websites of many counties, such as Australia, the US, Singapore, also have been tracking these numbers starting from various time points.   These websites have become  valuable resources to help advance the understanding of spread of the virus. We have access to  a time-series data repository on GitHub ( https://github.com/ulklc/covid19-timeseries), which consolidates and updates information obtained from these data sources. Our data analysis is based on the data obtained from this GitHub  data repository.
 } 
 
 Much effort has been devoted by the affected countries to battling the disease.  However, the crisis has not been over, with new infections detected every day. To forecast when the  pandemic gets controlled and evaluate the effects of virus control measures, it is imperative to develop appropriate  models to describe and understand the change trend of the pandemic \cite{hall2007real,grassly2008mathematical,chang2020modelling,pellis2020challenges}. 

    The susceptible-infectious-removed (SIR) model  was utilized to explain the rapid rise and fall of the infected individuals from the epidemics of  severe acute respiratory syndrome (SARS), {influenza A virus subtype} (H1N1) and middle east respiratory syndrome (MERS) \cite{laguzet2015individual,schwartz2015estimating,huang2016bayesian,mkhatshwa2010modeling,giraldo2008deterministic}. The key idea  is to divide a total population into  three  compartments: the susceptible, $S$, who are healthy individuals capable of contracting the disease; the {infectious}, $I$, who have the disease and are infectious; and the removed, $R$, who have recovered from the disease and gained immunity or who have died from the disease \cite{blackwood2018introduction}.  The model assumes a one-way flow  from susceptible to infectious to removed, and is reasonable for infectious diseases, which are transmitted from human to human, and where recovery confers lasting resistance \cite{cox2012risk}. {SIR models originated from  the Kermack-McKendrick model\cite{kermack1927contribution}, consisting of  three coupled differential  equations}  to describe the dynamics of the numbers in the $S, I $ and $R$ compartments, which tend to fluctuate over time. For example,  the number of infectious individuals increases drastically at the start of the epidemic, with a surge in susceptible individuals becoming infectious. As the epidemic develops, the number of infectious individuals  decreases when more infectious individuals die or recover than susceptible individuals become infectious. The epidemic ends when the infectious compartment ceases to exist \cite{kermack1927contribution,blackwood2018introduction}.

   SIR models and {the modified versions, such as susceptible-exposed-infectious-recovered model (SEIR),} were  applied to analyze the COVID-19 outbreak  \cite{Nesteruk2020,Chen2020,peng2020,zhou2020,Maier2020}.
Many of these models assume constant transmission and removal rates, which may not hold in reality.  For example, as a result of
various virus containment strategies,  such as  self-quarantine and  social distancing mandates, the transmission and removal rates may vary over time \cite{tognotti2013lessons}.

Recently, {a number of researchers \cite{timesir2020,sirmodel2018,epidemiological2020}} considered
time-dependent SIR models adapted to the  {dynamical epidemiological processes  evolving over time}.
However, few considered random errors in reporting, such as under-reporting (e.g. asymptomatic cases or {virus mutation}) or over-reporting (e.g.  false positives of testing), or characterized the uncertainty of  predictions.

     {Poisson models  naturally fit count data \cite{hilbe2014modeling}. Several works \cite{kirkeby2018methods,o2014fitting, zhuang2013multi} used Poisson distributions to model $I$ and $R$ from frequentist or Bayesian perspectives; however, most of the works only considered  constant transmission and removal rates. How to extend these works to accommodate time-dependent rates remains elusive. }

  We propose to adopt a Poisson model to estimate the time-varying transmission and removal rates, and understand the trends of the pandemic across countries. For example, we can  predict the number of the infectious persons and the number of removed persons at a certain time for each country, and forecast when the curves of cases become flattened. 
  
  An important {epidemiological} index that characterizes the transmission potential  is the basic reproduction number, ${\cal R}_0$,  defined as the expected number of secondary cases produced by an infectious case \cite{doi:10.1177/096228029300200103,diekmann2000mathematical,eichner2003transmission}. 
  {Our model leads to a temporally dynamical ${\cal R}_0$, }
  which measures at a given time how many people one infectious person, during the infectious period, will infect \cite{liu2018measurability}. This may help  evaluate the quarantine policies implemented by various authorities. A recent work \cite{liu2018measurability} demonstrated that ${\cal R}_0$ is likely to vary ``due to the impact of the performed intervention strategies and behavioral changes in the population."

The merits of our work  are summarized as follows.
First,  unlike  the deterministic ODE-based SIR models, our method does not require transmission and removal rates to be known, but estimates them using the data. 
 Second,  we allow these rates to be time-varying. Some time-varying SIR approaches \cite{epidemiological2020} directly integrate into the model the information on  when governments enforced, for example, quarantine, social-distancing, compulsory mask-wearing and city lockdowns.  Our method differs by computing a time-varying ${\cal R}_0$, which gauges the status of coronavirus containment and  assesses the effectiveness of virus control strategies. Third,   our Poisson model accounts for possible random errors in reporting, and quantifies the uncertainty of the predicted numbers of susceptible, infectious and removed. {Finally, we apply 
our method to analyze the data collected from the aforementioned GitHub time-series data repository. We have created an interactive web application (\url{https://younghhk.shinyapps.io/tvSIRforCOVID19/}) to facilitate users' application of the proposed method.}
 
\section{A Poisson model with  time-dependent transmission and removal rates}

  We introduce a Poisson model with  time-varying transmission and removal rates, denoted by  $\beta(t)$ and  $\gamma(t)$.  Consider a population with  $N$ individuals, and denote by  $S(t), I(t), R(t)$ the true but unknown numbers of  susceptible, infectious and  removed, respectively, at time $t$, and   by  $s(t) = S(t)/N$, $i(t) = I(t)/N$, $r(t) = R(t)/N$  the  fractions of  these  compartments. 
  
  \subsection{Time-varying transmission, removal rates and  reproduction number}
  The following  ordinary differential equations (ODE) describe the change rates of  $s(t), i(t)$ and  $r(t)$:
\begin{eqnarray}
\label{eq:11}
    \frac{ds(t)}{dt} &= & -\beta (t) s (t) i(t),   \\ \frac{di(t)}{dt} &= & \beta (t) s(t) i(t) -\gamma(t)  i(t), \label{eq:12}
    \\  \frac{dr(t)}{dt}& = & \gamma(t) i(t),  \label{eq:13}
\end{eqnarray} 
with an initial condition: $i(0)=i_0$ and $r(0)=r_0$, where $i_0>0$ in order to let the epidemic develop \cite{chen2020time}.  Here, $\beta(t)>0$  is  the time-varying transmission rate of an infection at time $t$, which is the number of infectious contacts  that result in infections per unit time,
and $\gamma(t)>0$ is the time-varying removal rate at $t$, at which infectious subjects are removed from being infectious due to death or recovery  \cite{diekmann2000mathematical}. Moreover, $\gamma^{-1}(t)$ can be interpreted as the infectious duration of an infection
caught at time $t$ \cite{notesonr0}.

{From (\ref{eq:11})-(\ref{eq:13})}, we derive an important quantity, which is the {time-dependent} reproduction number
\[
{\cal R}_0(t)= \frac{\beta(t)} {\gamma(t)}.
\]
 
To see this, {dividing (\ref{eq:12}) by  (\ref{eq:13})} leads to
\begin{equation} \label{change}
{\cal R}_0(t) = \frac{1}{s(t)} \left\{ \frac{di}{dr} (t) +1 \right\},
\end{equation}
where {$({di}/{dr})(t)$} 
is the ratio of the change rate of $i(t)$ to that of $r(t)$.
Therefore,  compared to its time-independent counterpart, ${\cal R}_0(t)$
 {is an instantaneous reproduction number and   provides a  real-time picture of an outbreak.} 
  For example, at the onset of the outbreak and in the absence of any containment actions, we may see a rapid ramp-up of cases compared to those removed, leading to  a large {$({di}/{dr}) (t)$} in (\ref{change}), and hence a large ${\cal R}_0(t)$. With the implemented policies for disease mitigation, we will see a drastically decreasing $({di}/{dr})(t)$ and, therefore, declining of ${\cal R}_0(t)$ over time.  The turning point is $t_0$ such that ${\cal R}_0(t_0)=1,$ {when the outbreak is controlled with $({di}/{dr})(t_0) < 0$.}

 {Under the fixed population size assumption, i.e., $s(t)+ i(t)+r(t)=1$, we only need to study $i(t)$ and $r(t)$, and } re-express (\ref{eq:11})-(\ref{eq:13}) as 
 \begin{eqnarray}
\label{eq:time}
 \frac{di(t)}{dt} & = & \beta(t) i(t)  \{1-i(t)-r(t)\} -\gamma(t) i(t), \nonumber \\  \frac{dr(t)}{dt} & = & \gamma(t) i(t), 
\end{eqnarray} 
with the same initial condition.  

\subsection{A Poisson model based on discrete  time-varying SIR}
As the numbers of cases and removed are reported on a daily basis,   $t$ is measured in days, e.g.  $t=1,\ldots,T$. Replacing  derivatives in \eqref{eq:time} with finite differences, we can consider a discrete version of {\eqref{eq:time}: 
\begin{eqnarray} \label{discrete}
i(t+1)-i(t) & = & \beta(t) i(t) \{1-i(t)-r(t)\} -\gamma(t) i(t),  \nonumber \\
r(t+1)-r(t) & = & \gamma(t) i(t), \end{eqnarray}
where $\beta(t)$ and $\gamma(t)$ are positive functions of $t$. We set
$i(0)=i_0$ and $r(0)=r_0$
with $t=0$ being the starting date.

 Model (\ref{discrete}) admits a recursive way to compute $i(t)$ and $r(t)$:
\begin{eqnarray} \label{solution}
 i(t+1) &= & \{1+ \beta(t)-\gamma(t) \} i(t) - \beta(t) i(t)\{i(t)+r(t)\},  \nonumber \\
r(t+1) & = &  r(t) + \gamma(t) i(t) 
\end{eqnarray}
for $t=0, \ldots, T-1$.  The first equation of  (\ref{solution}) implies that  $\beta(t) < \gamma(t)$  or $ {\cal R}_0(t)=\beta (t)\gamma^{-1}(t) <1$ leads to that $i(t+1)< i(t)$ or the number of infectious cases drops, meaning the spread of virus is controlled; otherwise, the number of infectious cases will keep increasing. }

\subsection{Estimation and inference}

{{To fit the model and estimate the time-dependent parameters, we can use nonparametric techniques, such as splines {\cite{de1978practical,ramsay2007parameter,lawrence2007modelling,alvarez2013linear,alvarez2009latent,wheeler2014mechanistic}}, local polynomial regression \cite{ruppert1997empirical} and reproducible kernel Hilbert space method \cite{akgul2015new}. 
In particular, we consider a cubic B-spline approximation \cite{perperoglou2019review}.  

Denote by $\bB(t) =\{B_1(t), \ldots, B_q(t)\}^\rT $ the $q$ cubic B-spline basis functions over $[0,T]$   associated with the knots $0 = w_0 < w_1 < \dots < w_{q-2} < w_{q-1} = T$.
{For added flexibility, we allow the number of knots to differ between $\beta(t)$ and $\gamma(t)$ and}
specify  
{
\begin{eqnarray} \label{poly}
\log \beta(t) & = &   \sum_{j=1}^{q_1} \beta_j B_j(t),  \nonumber \\
\log \gamma(t) & = &   \sum_{j=1}^{q_2} \gamma_j B_j(t).
\end{eqnarray}
When $\beta_1=\cdots =\beta_{q_1}$
and $\gamma_1=\cdots =\gamma_{q_2}$,} the model reduces to a constant SIR model \cite{perperoglou2019review}. }
We use cross-validation to choose $q_1$ and $q_2$ in our numerical experiments. 

Denote by {$\bbeta = (\beta_1, \ldots, \beta_{q_1})$} and  {$ \bgamma = (\gamma_1, \ldots, \gamma_{q_2})$} the unknown parameters,
by  $Z_I(t)$ and $Z_R(t)$ the reported numbers of infectious and removed, respectively, and by $z_I(t)=Z_I(t)/N$ and $z_R(t)=Z_R(t)/N$,
the reported proportions.   
Also, denote by $I(t)$ and $R(t)$ the true numbers of infectious and removed, respectively at time $t$.  We propose a Poisson model to link $Z_I(t)$ and $Z_R(t)$ to $I(t)$ and $R(t)$ as follows:   
\begin{eqnarray} \label{pois}
Z_R(t) & \sim &  \text{Pois} \{R(t)\},   \nonumber \\
Z_I(t) &  \sim & \text{Pois} \{I(t)\}.
\end{eqnarray}
We also assume that,  given $I(t)$ and $R(t)$, 
the {observed} daily number $\{Z_I(t), Z_R(t)\}$ are independent across $t=1, \ldots, T$, meaning the random reporting errors are ``white" noise. {We note  that  (\ref{pois}) is directly based on ``true" numbers of infectious cases and removed cases derived from the discrete SIR model (\ref{discrete}). This differs from the Markov process approach, which is based on the past observations.}

With  (\ref{discrete}),  (\ref{solution}) and  (\ref{poly}),   $R(t)$ and $I(t)$ are the functions of  $\bbeta$ and $\bgamma$,  since  $ R(t)=N \times r(t)$  and $I(t)=N \times i(t)$.  Given  the data $(Z_I(t), Z_R(t)), t= 1, \ldots, T$, 
we  obtain $(\hat{\bbeta},\hat{\bgamma})$, the estimates of $(\bbeta,\bgamma)$, by maximizing the {following}  likelihood
\[L(\bbeta,\bgamma)=\prod_{t=1}^{T}\frac{e^{-R(t)}R(t)^{Z_{R(t)}}}{ Z_{R}(t)!}\times\prod_{t=1}^{T}\frac{e^{-I(t)}I(t)^{Z_I(t)}}{Z_I(t)!},\]
or, equivalently, maximizing the 
log likelihood function
\begin{equation}\label{loglik} 
\ell(\bbeta,\bgamma)= N \sum_{t=1}^T \left\{ -r(t) + z_R(t) \log r(t) - i(t) + z_I(t) \log i(t)  \right\} + C,
\end{equation}
where $C$ is a constant free of $\bbeta$ and $\bgamma$. See the Appendix for additional details of {optimization}.

We then estimate the variance-covariance matrix of $(\hat{\bbeta},\hat{\bgamma})$ by inverting the second derivative of $-\ell(\bbeta,\bgamma)$ evaluated at $(\hat{\bbeta},\hat{\bgamma})$.
{Finally,  for $t=1, \ldots, T$, we estimate $I(t)$ and $R(t)$  by  $\hat{I}(t) = N \hat{i}(t)$ and $\hat{R}(t) = N \hat{r}(t)$,
where   $\hat{i}(t)$ and  $\hat{r}(t)$ are obtained from
(\ref{solution}) with all unknown quantities replaced by their estimates;}
{ estimate $\beta(t)$ and $\gamma(t)$  by $\hat{\beta}(t)$ and $\hat{\gamma}(t)$, obtained by using
(\ref{poly}) with  $({\bbeta},{\bgamma})$  replaced by  $(\hat{\bbeta},\hat{\bgamma})$; and
estimate ${\cal R}_0(t)$ by $\hat{{\cal R}}_0(t) = \hat{\beta}(t)/ \hat{\gamma}(t)$.}

\vspace{0.25in}
\hrule \vspace{0.05in}
{\bf Summary of estimation and inference for $\beta(t)$, $\gamma(t)$}, ${\cal R}_0(t)$,
$I(t), R(t)$
\vspace{0.05in}
\hrule
\begin{enumerate}[]
\item {\bf  Estimation:} 
Let $N$  be the size of population of a given country.  The date when the first case was reported is set to be the starting date with $t=1$, $i_0= Z_I(1)/N$ and $r_0= Z_R(1)/N$.  The observed data are
 $\{Z_I(t),Z_R(t), t=1,\ldots, T\}$, {obtained from the  GitHub data repository website mentioned in the introduction.}
{We maximize (\ref{loglik}) to obtain $\hat{\bbeta}=(\hat{\beta}_0,\hat{\beta}_1,\ldots,\hat{\beta}_{q_1})$ and $\hat{\bgamma}=(\hat{\gamma}_0,\hat{\gamma}_1,\ldots,\hat{\gamma}_{q_2})$. The optimal $q_1$ and $q_2$ are obtained via cross-validation.  We  denote by 
$\hat{\bbeta}=(\hat{\beta}_0,\hat{\beta}_1,\ldots,\hat{\beta}_{q_1})$ and $\hat{\bgamma}=(\hat{\gamma}_0,\hat{\gamma}_1,\ldots,\hat{\gamma}_{q_2})$, based on which  we calculate  $\hat{\beta}(t), \hat{\gamma}(t), \hat {\cal R}_0(t), \hat{R}(t),  \hat I(t)$.}

\item  {\bf  Inference:}  The estimated variance-covariance matrix of $(\hat{\bbeta},\hat{\bgamma})$, denoted by $\hat{V}(\hat{\bbeta},\hat{\bgamma})$, can be obtained by inverting the second derivative of $-\ell(\bbeta,\bgamma)$ evaluated at $(\hat{\bbeta},\hat{\bgamma})$. For each $t$, as  $\hat{\beta}(t)$, $\hat{\gamma}(t)$, $\hat {\cal R}_0(t)$, $\hat{R}(t)$ and  $\hat I(t)$
are smooth
functions of $\hat{\bbeta}$ and $\hat{\gamma}$, we apply  the delta method \cite{parr1983note}
 to estimate
their variances  and obtain the confidence intervals. As an illustration, we  compute
$\widehat{\mbox{var}} (\hat R(t))  = \dot{\hat{R}}(t)^\rT \hat{V}(\hat{\bbeta},\hat{\bgamma}) \dot{\hat{R}}(t)$
and 
$\widehat{\mbox{var}} (\hat I(t))  = \dot{\hat{I}}(t)^\rT \hat{V}(\hat{\bbeta},\hat{\bgamma}) \dot{\hat{I}}(t), $ where $\dot{\hat{R}}(t)$   and
$\dot{\hat{I}}(t)$ are the partial derivative vectors of $\hat{R}(t)$ and $\hat{I}(t)$ with respect to $(\hat{\bbeta},\hat{\bgamma})$.

\end{enumerate}
\hrule \vspace{0.1in}

 \section{Analysis of the COVID-19 pandemic among severely affected countries} 

Since the first case of COVID-19 was detected in  China,  it quickly spread to nearly every part of the world \cite{jhu}. COVID-19, conjectured to be more contagious than the previous SARS and H1N1
\cite{sansonetti2020covid},  has put great strain on  healthcare systems worldwide, especially among the severely affected countries \cite{tanne2020covid}. We apply  our method to assess the epidemiological processes of COVID-19 in some  severely impacted countries.   

\subsection{\bf Data descriptions and robustness of the method towards specifications of the initial conditions}

{The country-specific time-series data of confirmed, recovered, and death cases were obtained from a GitHub data repository website (https://github.com/ulklc/covid19-timeseries).  This site collects information from various sources listed below on a daily basis at GMT 0:00,  converts the data to the CSV format, and conducts data normalization and harmonization if inconsistencies are found. The data sources include    
\begin{itemize}
    \item  World Health Organization (WHO): https://www.who.int/
    \item DXY.cn. Pneumonia 2020: http://3g.dxy.cn/newh5/view/pneumonia.
    \item BNO News: https://bnonews.com/index.php/2020/02/the-latest-coronavirus-cases/
    \item National Health Commission of China (NHC): \\
    http://www.nhc.gov.cn/xcs/yqtb/list\_gzbd.shtml
    \item China CDC (CCDC): http://weekly.chinacdc.cn/news/TrackingtheEpidemic.htm
    \item Hong Kong Department of Health: https://www.chp.gov.hk/en/features/102465.html
    \item Macau Government: https://www.ssm.gov.mo/portal/
    \item Taiwan CDC: https://sites.google.com/cdc.gov.tw/2019ncov/taiwan?authuser=0
    \item US CDC: https://www.cdc.gov/coronavirus/2019-ncov/index.html
    \item Government of Canada: \\ https://www.canada.ca/en/public-health/services/diseases/coronavirus.html
    \item Australia Government Department of Health: \\ https://www.health.gov.au/news/coronavirus-update-at-a-glance
    \item European Centre for Disease Prevention and Control (ECDC): \\ https://www.ecdc.europa.eu/en/geographical-distribution-2019-ncov-cases
    \item Ministry of Health Singapore (MOH): https://www.moh.gov.sg/covid-19
    \item Italy Ministry of Health: http://www.salute.gov.it/nuovocoronavirus
    \item Johns Hopkins CSSE: https://github.com/CSSEGISandData/COVID-19
    \item WorldoMeter: https://www.worldometers.info/coronavirus/
    \end{itemize}}
    {In particular, the current population size of each country, $N$, came from the website of WorldoMeter.} 
{Our analyses covered the periods between the date of the first reported coronavirus case in each nation and {June 30, 2020.}}
{In the beginning of the outbreak,   assessment of $i_0$ and $r_0$ was problematic as  infectious but asymptomatic cases tended to be  undetected due to  lack of awareness and testing. To investigate how our method depends on the correct specification of the  initial values  $r_0$  and $i_0$, we conducted Monte Carlo simulations. As a comparison, we also studied the performance of the deterministic SIR model in the same settings.   Figure~\ref{fig:initial} shows that, when the  initial value $i_0$ was mis-specified to be  5 
times of the truth, the curves of $i(t)$ and $r(t)$ obtained by the  deterministic  SIR model (\ref{discrete}) were considerably biased.  On the other hand, our  proposed model (\ref{pois}), by accounting for the randomness of the observed data,  was  robust toward  the mis-specification of  $i_0$ and $r_0$: the  estimates of $r(t)$ and $i(t)$ had negligible biases even with mis-specified initial values.  In an omitted analysis, we  mis-specified $i_0$ and $r_0$ to be only twice of the truth, and obtain the {similar} results.

Our numerical experiments also  suggested that using the time series, starting from the date when both cases and removed were reported, may generate more reasonable estimates.}

%

\subsection{Estimation of country-specific transmission, removal rates and reproduction numbers}

{Using the cubic B-splines (\ref{poly}),
we estimated the time-dependent transmission rate $\beta(t)$ and removal rate $\gamma(t)$, based on which we further estimated ${\cal R}_0(t)$, $I(t)$  and $R(t)$.  To choose  the optimal number of  knots for each country when implementing the spline approach,  we used 5-fold cross-validation by  minimizing the combined mean squared  error for the estimated infectious and removed cases.}

   {Figure~\ref{fig:us} shows sharp variations in transmission rates and removal rates across different time periods, indicating the time-varying nature of these rates.}  The estimated $I(t)$ and $R(t)$   overlapped well with  the observed number of infectious  and removed cases, indicating the reasonableness of the method. The pointwise 95\% confidence intervals (in yellow) represent the uncertainty of the estimates, which may be due to error in reporting.
Figure~\ref{fig:all} presents the estimated time-varying reproduction number, $\hat{\beta}(t)\hat{\gamma}(t)^{-1}$,  for several countries. The curves capture the evolving  trends of the epidemic for each country.  

In the US, though the first confirmed case  was reported on January 20, 2020,  lack of immediate actions in the early stage let the epidemic spread widely. As a result, the US had seen soaring infectious cases, and ${\cal R}_0(t)$ reached  its peak  around mid-March. From mid-March to  early April, the US tightened the virus control policy by suspending foreign travels and closing borders, and the federal government and most states issued mandatory or advisory stay-home orders, which seemed to have substantially contained the virus.  

The  high reproduction numbers with China, Italy, and Sweden at
the onset of the pandemic imply that the spread of the infectious disease was not well controlled in its early phases.  With the extremely stringent mitigation policies such as city lockdown and mandatory mask-wearing implemented in the end of January, China was reported to bring its epidemic under control with a quickly dropping ${\cal R}_0(t)$ in February. This indicates that China might have contained the epidemic, with more people removed from infectious status than those who became infectious.
  
{Sweden is among the few countries that imposed more relaxed measures to control coronavirus  and advocated herd immunity. The Swedish approach has initiated much debate. While some criticized that this may endanger the general population in a  reckless way,  some felt this might terminate the pandemic more effectively in the absence of vaccines \cite{sweden}. Figure~\ref{fig:all} demonstrates that Sweden has a large reproduction number, which however keeps decreasing. The ``big V" shape of the reproduction number around May 1  might be due to the reporting errors or lags. Our investigation found that the reported number of infectious cases in that period suddenly dropped and then quickly rose back, which was unusual.}}

  Around February 18, a  surge in South Korea was linked to  a massive cluster of more than 5,000 cases \cite{wsj2020}. The outbreak was  clearly depicted in the time-varying
 ${\cal R}_0(t)$ curve. Since then, South Korea  appeared to have  slowed its epidemic, likely due to expansive  testing programs and extensive efforts to  trace and isolate patients and their contacts \cite{science2020}.


{More broadly, Figure \ref{fig:all} categorizes countries into two groups. One group features the countries which have contained coronavirus.  Countries, such as China and South Korea, took aggressive actions after the outbreak and presented sharper downward slopes. Some European countries such as  Italy and Spain and  {Mideastern countries} such as Iran,  which were hit later than the East Asian countries, share a similar pattern, though with much {flatter} slopes.  On the other hand,  the US,   Brazil, and Sweden are  still struggling to contain the virus, with  the ${\cal R}_0(t)$ curves  hovering over 1.} We also caution that, {among  the countries whose ${\cal R}_0(t)$ dropped below 1,}   the curves of the reproduction numbers are beginning to uptick,  possibly due to the resumed economy activities.

\subsection{An interactive web application and R code}

\begin{figure}[!htbp]
	\centering
	\includegraphics[width=5in]{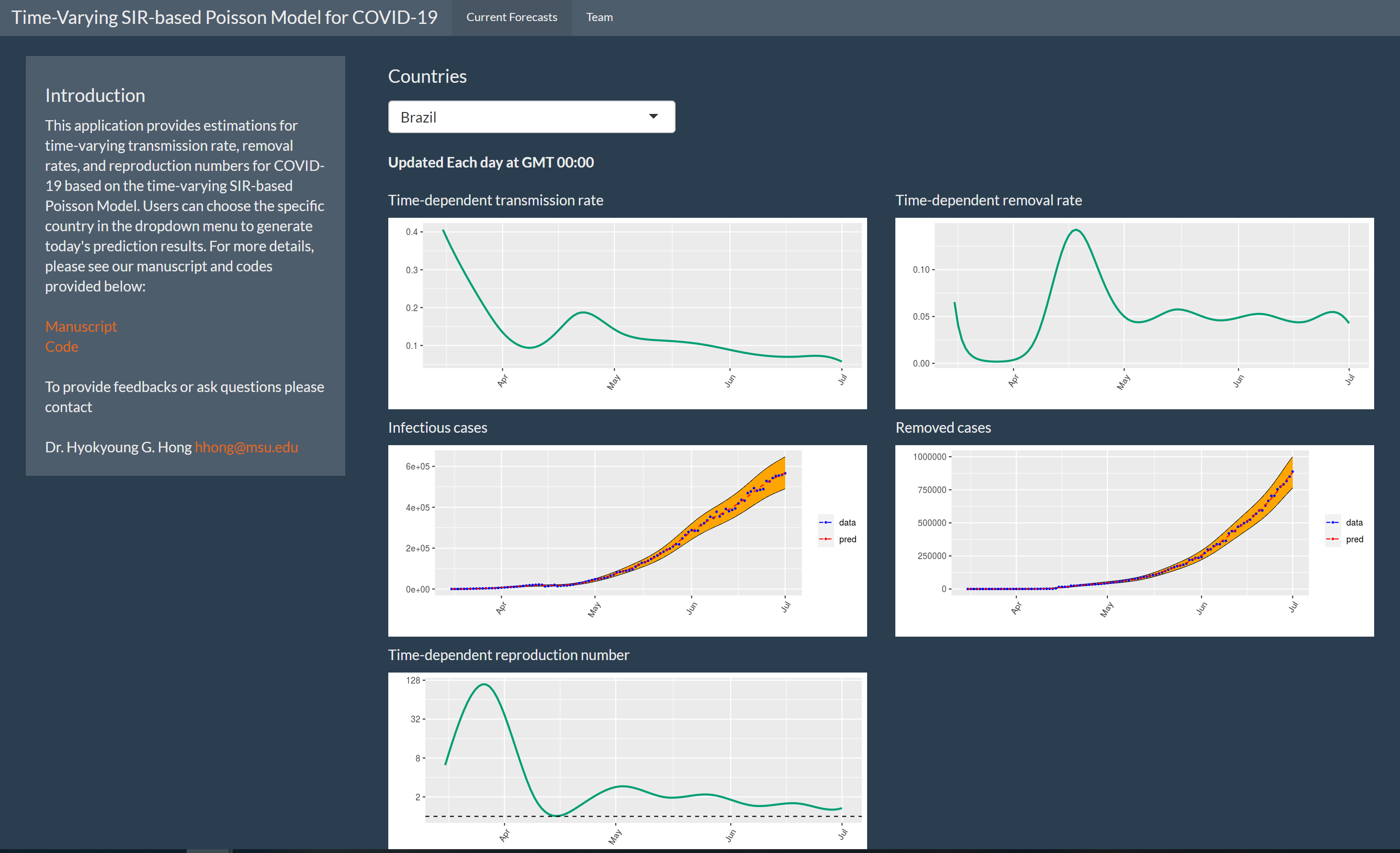}
	\caption{An illustration of the developed interactive web application.}
	\label{fig:shiny}
\end{figure}

We have developed a web application
(\url{https://younghhk.shinyapps.io/tvSIRforCOVID19/}) to facilitate users' application of the proposed method  to compute  the time-varying  reproduction number, and estimated and predict the daily numbers of active cases and removed cases for the presented countries and other countries; see Figure \ref{fig:shiny} for an illustration.

{Our code  was written in \texttt{R} \cite{citeR}, using the \texttt{bs} function in the \texttt{splines} package for cubic B-spline approximation,  the \texttt{nlm} function in the \texttt{stats} package for nonlinear minimization, and the \texttt{jacobian}
 function in the \texttt{numDeriv} package for computation of gradients and hessian matrices. Graphs were made by using the \texttt{ggplot2} package. Our code can be found on the aforementioned shiny website.}


\section{Discussion}

The rampaging  pandemic of COVID-19  has  called for developing proper computational and statistical tools to understand the trend of the spread of the disease and evaluate the efficacy of mitigation measures \cite{hossain2020current,sarkodie2020investigating,zhang2020impact, picchiotti2020covid}.
   We propose a Poisson  model with time-dependent     transmission and removal rates. Our  model   accommodates  possible random errors  and   estimates a time-dependent disease  reproduction number, ${\cal R}_0(t)$, which can serve as a metric for timely evaluating the effects of health policies.

    {There have been substantial issues, such as biases and lags,  in reporting infectious cases, recovery, and deaths,  especially at the early stage of the outbreak. As opposed to the deterministic SIR models that heavily rely on accurate reporting of initial infectious and removed cases,  our model is more robust towards  mis-specifications of such initial conditions.}
   Applications of our method to study the epidemics in selected countries  illustrate the results of the virus containment policies implemented in these countries, and may serve as the epidemiological benchmarks for the future preventive  measures.
  
  Several methodological questions need to be addressed. First, we analyzed each country separately, without considering the traffic flows among these countries. We will develop a joint model for the global epidemic, which accounts for the geographic locations of 
 and the connectivity among the countries.

  {Second, incorporating timing of public health interventions such as the shelter-in-place order  into the model might be interesting. However, we opted not to follow this approach as no such information exists for the majority countries. On the other hand, the impact of the interventions or the change point can be embedded into our nonparametric time-dependent estimates.}

{Third, the validity of the results of statistical models eventually hinges on the data transparency and accuracy. 
For example, the results of Chinazzi et al. \cite{chinazzi2020effect} suggested} that in China only one of four cases were detected and confirmed. Also,  asymptomatic cases might have been undetected in many countries.  All of these might have led to underestimation of the actual number of cases.
Moreover, the collected data could be biased toward patients with severe  infection and with insurance, as  these patients were more likely to seek care or get tested. More in-depth research is warranted to address the issue selection bias.

  {Finally, our present work is within  the SIR  framework, {where removed individuals include recovery and deaths, who hypothetically are unlikely to infect others.} Although this makes the model simpler and widely adopted, the interpretation of the $ \gamma$ parameter is not straightforward. Our subsequent work is to develop a susceptible-infectious-recovered-deceased  (SIRD) model, in which the number of deaths and the number of recovered are separately considered. We will report this elsewhere.}

  \section{Conclusion}

  Containment of COVID-19 requires the concerted effort of health care workers, health policy makers as well as citizens.  Measures, e.g. self-quarantine,  social distancing, and shelter in place,  have been executed at various phases by each country to prevent the community transmission. Timely and effective assessment of these actions constitutes a critical component of the effort. SIR models have been widely used to model this pandemic. However, constant transmission and removal rates may not capture the timely influences of these policies.
  
We propose a time-varying SIR Poisson model to assess the dynamic transmission patterns of  COVID-19. 
{With  the virus containment measures taken at various time points,  ${\cal R}_0$ may vary substantially over time. Our model provides a systematic and daily updatable tool to evaluate the immediate outcomes of these actions.} 
It is likely that the pandemic is ending and many countries are now shifting  gear  to reopen the economy, while preparing to  battle {the second wave of virus attack}\cite{chinasecond,bbc}. Our tool may  shed light on and aid the implementation of future  containment strategies.



{\section*{Appendix: Details of Optimization}

To minimize (\ref{loglik}), we differentiate $\ell(\bbeta,\bgamma)$ with respect to $(\bbeta,\bgamma)$. Then  $(\hat{\bbeta},\hat{\bgamma})$ solves the following estimating equations: 

\begin{eqnarray*} \label{est-eq} 
\sum_{t=1}^T \left[ \left\{ \frac{z_R(t)}{ r(t)} -1 \right\} 
\frac{\partial r(t)} {\partial \bfbeta} + \left\{ \frac{z_I(t)}{ i(t)} -1 \right\}  
\frac{\partial i(t)} {\partial \bfbeta} \right] &=& 0,  \nonumber     \\
 \sum_{t=1}^T \left[ \left\{ \frac{z_R(t)}{ r(t)} -1 \right\} 
\frac{\partial r(t)} {\partial \bgamma} + \left\{ \frac{z_I(t)}{ i(t)} -1 \right\}  
\frac{\partial {i(t)}} {\partial \bgamma}\right] &=& 0,
\end{eqnarray*}
where the involved partial derivatives can be computed recursively. Specifically, taking partial derivatives on the both sides of (\ref{solution}) yields that, 
for $t=0, \ldots, T-1$, 
\begin{eqnarray*} 
\frac{\partial {i(t+1)} } {\partial \bbeta}  & = & [ 
1+\beta(t) -\gamma(t) -2 \beta(t) i(t) - \beta(t) r(t)]
\frac{\partial i(t) } {\partial \bbeta}
+  [ i(t) -i^2(t) -i(t)r(t)] \frac{\partial \beta(t)  } {\partial \bbeta} -\beta(t)i(t) \frac{\partial r(t)} {\partial \bbeta},   \\ 
\frac{\partial i(t+1)} {\partial \bgamma} & = & 
[ 1+\beta(t)- \gamma(t) - 2 \beta(t) i(t)-\beta(t)r(t)]
\frac{\partial i(t) } {\partial \bgamma}
- i(t) \frac{\partial \gamma(t) } {\partial \bgamma} -\beta(t)i(t) \frac{\partial r(t) } {\partial \bgamma}, \\
\frac{\partial r(t+1)} {\partial \bbeta} & = & 
\frac{\partial r(t)} {\partial \bbeta} + \gamma(t) \frac{\partial i(t)} {\partial \bbeta}, \\
\frac{\partial r(t+1)} {\partial \bgamma} & = & 
\frac{\partial r(t)} {\partial \bgamma} +  \gamma(t)\frac{\partial i(t)} {\partial \bgamma}  + i(t) \frac{\partial \gamma(t) } {\partial \bgamma}.
\end{eqnarray*}
Here,
$\frac{\partial} {\partial \bbeta}{\beta(t)}
= \beta(t) \times \bB(t)$, and
$\frac{\partial} {\partial \bgamma}{\gamma(t)}
= \gamma(t) \times \bB(t)$, and
$\frac{\partial} {\partial \bbeta}{i(0)}
= \frac{\partial} {\partial \bgamma}{i(0)}
=\frac{\partial} {\partial \bbeta}{r(0)}
= \frac{\partial} {\partial \bgamma}{r(0)}
=0$ by using the initial conditions.
}

\nolinenumbers
\bibliography{ref}

\begin{figure}[ht]
	\centering
	\includegraphics[width=3.5in]{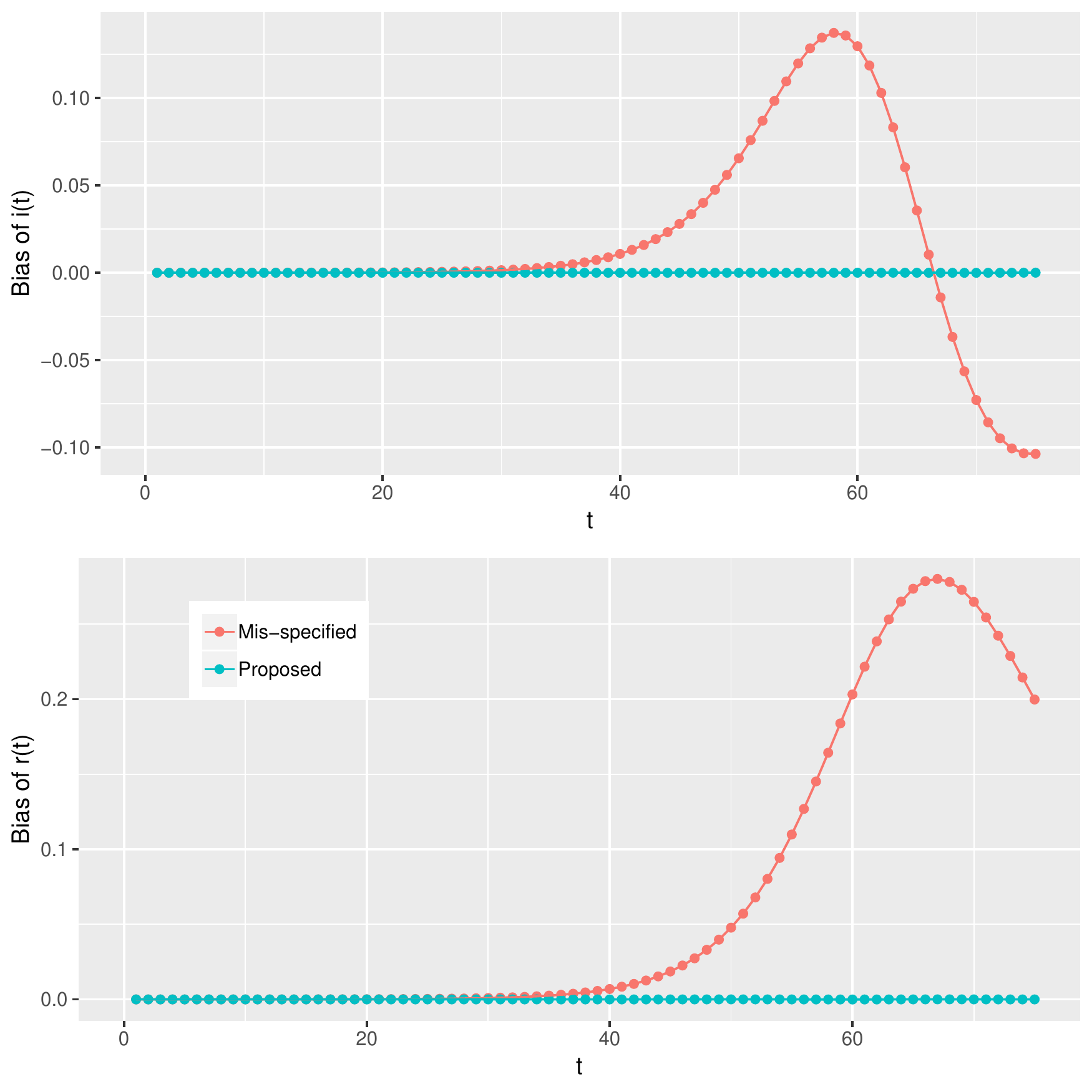}
	\caption{{
	{Plots of the relative biases of $\hat{i}(t)$  (upper) and  $\hat{r}(t)$  (lower)}   when $\hat{i}(t)$ and $\hat{r}(t)$ are derived  1) by using the ODE framework with the mis-specified initials  (``Mis-specified") and 2) by  using  proposed model  (\ref{pois})
	with  mis-specified initials   (``Proposed").
	In the model, the true  
	  $(\beta,\gamma)=(e^{-1},e^{-1.95})$ and $(i_0,r_0 )=(10^{-6},10^{-6})$ in (\ref{discrete}).} These values are roughly equal to the constant estimates of the real situation.	 The mis-specified initials are set as $(5\times10^{-6},5\times10^{-6})$.} 	\label{fig:initial}
\end{figure}

\begin{figure}[!htbp]
	\centering
	\includegraphics[width=6in]{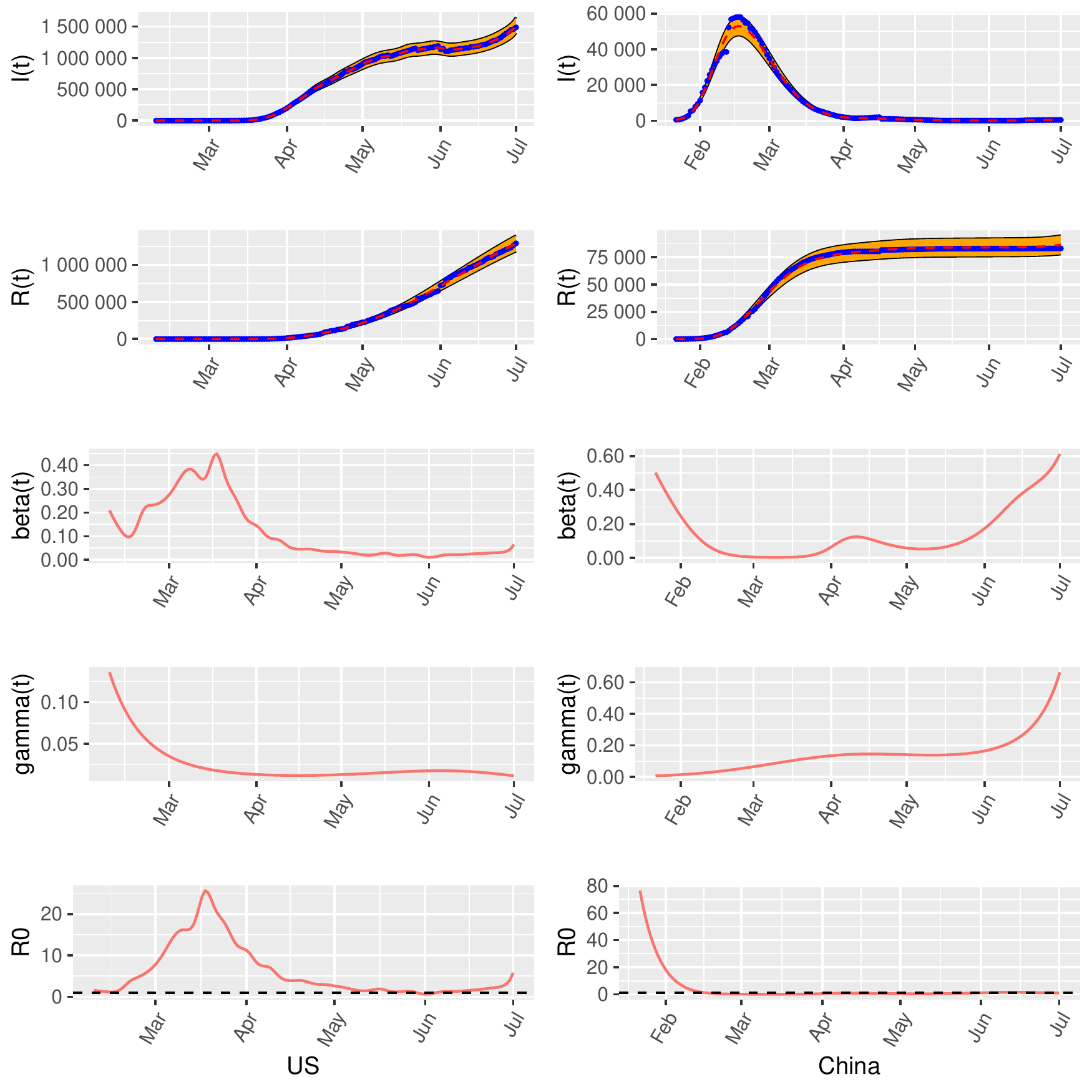}
	\caption{{Estimated $I(t)$, $R(t)$, $\beta(t)$, $\gamma(t)$,  and $\mathcal{R}_0(t)$ for the US (left) and China (right), based on the data up to {June 30, 2020}. The blue dots and the red dashed curves represent  the observed data and  the model-based predictions, respectively, with 95\% confidence interval. 
	}}
	\label{fig:us}
	\end{figure}
	
  \begin{figure}[!h]
	\centering
	\includegraphics[width=6in]{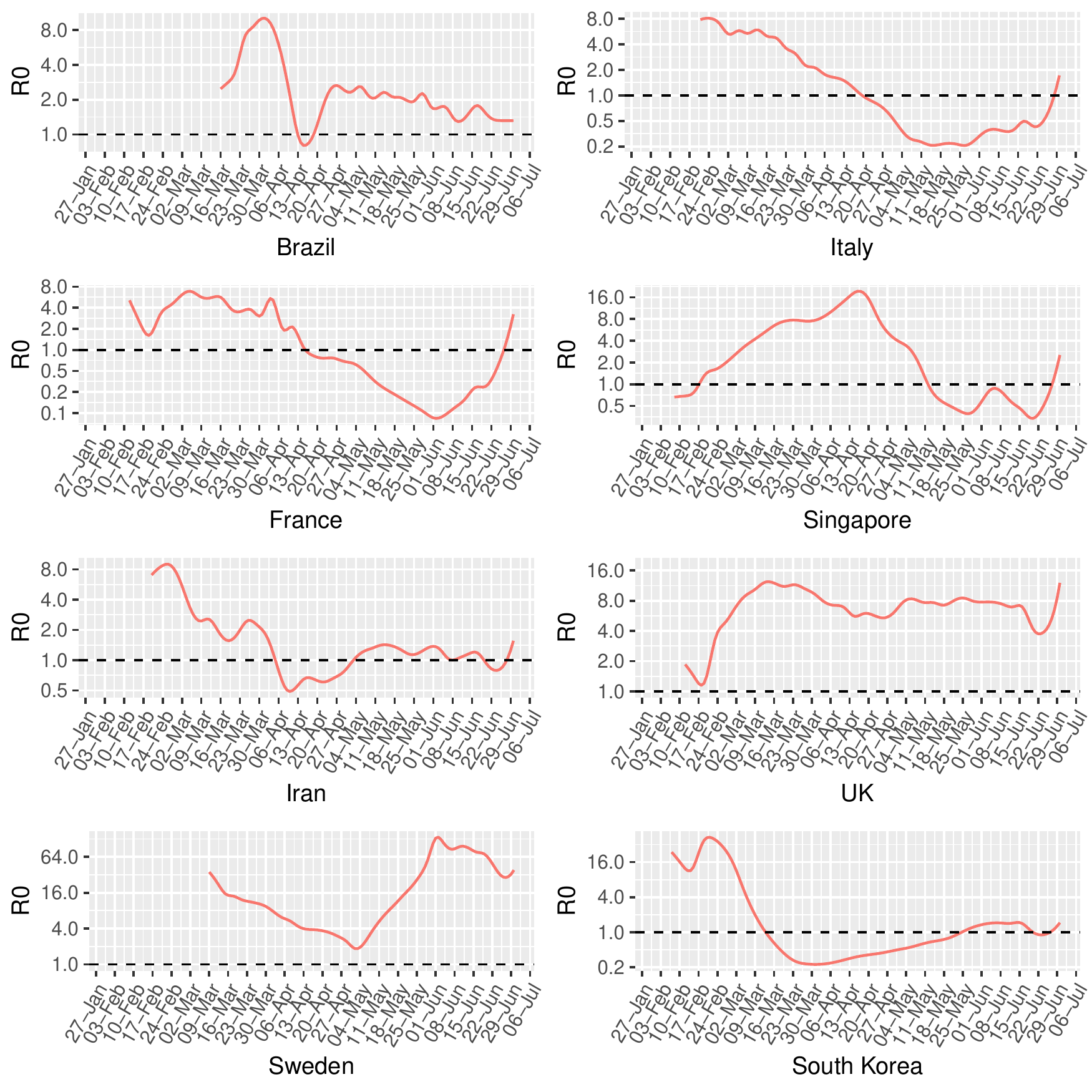}
	\caption{{Estimated  reproduction number ${\cal R}_0(t)$ for selected countries based on the data up to {June 30}, 2020.}}
	\label{fig:all}
\end{figure}

\end{document}